\documentclass[]{aa}
\usepackage{graphicx,natbib}

\def\jh{\mbox{$\rm (J-H)$}}

\def\jk{\mbox{$\rm (J-K_s)$}}

\def\ebv{\mbox{$\rm E(B-V)$}}
\def\ejh{\mbox{$\rm E(J-H)$}}
\def\rc{\mbox{$\rm R_{core}$}}
\def\rl{\mbox{$\rm R_{RDP}$}}
\def\rt{\mbox{$\rm R_{t}$}}
\def\ms{\mbox{$\rm M_\odot$}}
\def\ds{\mbox{$\rm d_\odot$}}
\def\tdis{\mbox{$\rm t_{dis}$}}

\def\Rgc{\mbox{$\rm R_\odot$}}
\def\dgc{\mbox{$\rm R_{GC}$}}
\def\xgc{\mbox{$\rm x_{GC}$}}
\def\ygc{\mbox{$\rm y_{GC}$}}
\def\zgc{\mbox{$\rm z_{GC}$}}

\def\jj{\mbox{$\rm J$}}
\def\hh{\mbox{$\rm H$}}
\def\ks{\mbox{$\rm K_s$}}
\def\aV{\mbox{$\rm A_V$}}
\def\ns{\mbox{$\rm N_{1\sigma}$}}
\def\no{\mbox{$\rm N_{obs}$}}
\def\nc{\mbox{$\rm N_{cl}$}}
\def\sFS{\mbox{$\rm\sigma_{FS}$}}
\def\fsU{\mbox{$\rm FS_{unif}$}}

\begin{document}

\title{Exploring FSR open cluster candidates within $|\Delta\ell|=20^\circ$
of the Galactic anticentre}

\author{C. Bonatto\inst{1} \and E. Bica\inst{1}}

\offprints{C. Bonatto}

\institute{Universidade Federal do Rio Grande do Sul, Departamento de Astronomia\\
CP\,15051, RS, Porto Alegre 91501-970, Brazil\\
\email{charles@if.ufrgs.br, bica@if.ufrgs.br}
\mail{charles@if.ufrgs.br} }

\date{Received --; accepted --}

\abstract
{We investigate the nature of a sample of star cluster candidates detected as stellar 
overdensities towards the Galactic anticentre.}
{Taken from the catalogue of Froebrich, Scholz, and Raftery (FSR), the sample contains 
28 star cluster candidates located within $|\Delta\ell|=20^\circ$ of the anticentre. 
These are all the candidates in that sector classified by FSR with a high probability
of being star clusters. Our main goals are to determine the fraction of such candidates that
are unknown star clusters, to derive their astrophysical parameters, and to investigate the
relationship of cluster parameters with position in the Galaxy.}
{Properties of the star cluster candidates are investigated with field-star decontaminated
2MASS colour-magnitude diagrams and stellar radial density profiles.}
{All candidates present significant excesses in the radial density profiles, consistent with
the method from which they were originally selected. Of the 28 candidates, 7 are previously 
known open clusters, 2 have been recently identified, and 6 are new ones with ages from 30\,Myr
to 1\,Gyr. Among the remaining 13 candidates, 6 are uncertain cases that require deeper observations,
while 7 appear to be important field fluctuations. The structure of part of the newly identified
open clusters appears to be affected by interaction with giant molecular clouds in the Local and 
Perseus arms.}
{When photometric and radial distribution properties are considered together, an important fraction 
of the stellar overdensities with a fluctuation level $\ga3\sigma$ are shown to be star clusters. 
Thus, catalogues of star cluster candidates, coupled to the present kind of study, are an important 
source for identifying unknown open clusters. Such efforts affect the understanding of the
star-formation rate, cluster dynamical evolution, and Galactic structure, among others.}


\titlerunning{FSR star cluster candidates towards the anticentre}

\authorrunning{C. Bonatto and E. Bica}

\maketitle

\section{Introduction}
\label{intro}

The open clusters (OCs) dwell and evolve in the Galactic disk.
Dynamical processes such as mass segregation and evaporation, tidal interactions with the
disk and bulge, and collisions with giant molecular clouds, as well as mass loss from the
stellar evolution, accelerate the dynamical evolution, which affects the cluster structure
in varying degrees. Given these circumstances, most OCs end up completely dissolved in the
Galactic stellar field (e.g. \citealt{Lamers05}) or as remnants (\citealt{PB07}
and references therein).

Evidence on different grounds, such as theoretical (e.g. \citealt{Spitzer58}; \citealt{LG06}), 
N-body (e.g. \citealt{BM03}; \citealt{GoBa06}; \citealt{Khalisi07}), and observational (e.g. 
\citealt{vdB57}; \citealt{Oort58}; \citealt{vHoerner58}; \citealt{Piskunov07}), consistently point to a
disruption-time (\tdis) scale that increases with Galactocentric distance. While in the inner Galaxy
massive clusters are dissolved in $\tdis\sim50$\,Myr (\citealt{Portegies02}), near the Solar circle
the disruption-time scale is shorter than $\sim1$\,Gyr (e.g. \citealt{Bergond2001}; \citealt{Lamers05}).
Reflecting this dependence, OCs older than $\sim1$\,Gyr are preferentially found near the Solar circle
and in the outer Galaxy (e.g. \citealt{vdBMc80}; \citealt{Friel95}; \citealt{DiskProp}).

According to the WEBDA\footnote{\em obswww.univie.ac.at/webda - \citet{Merm03}} database, the
current census provides $\sim1000$ OCs with known parameters. The statistics, however, are far from
complete, especially at the faint end of the luminosity distribution and large distances (e.g.
\citealt{Kharchenko05}; \citealt{Piskunov07}; \citealt{DiskProp}). Besides dynamical disruption,
observational limitations due to low cluster/background contrast, restrict the detectability to a very
small fraction of the OCs in the Galaxy (\citealt{DiskProp}). Thus, the derivation of astrophysical
parameters of unknown star clusters is an important step in defining their statistical properties better.

A catalogue of 1021 star cluster candidates for $|b|\le20^\circ$ and all Galactic longitudes was
published by \citet{FSRcat}. Based essentially on stellar number-densities, they identified
small-scale regions in the 2MASS\footnote{The Two Micron All Sky Survey --- {\em
www.ipac.caltech.edu/2mass/releases/allsky/ }} database as overdensities with respect to the
surroundings. The overdensities were classified according to a quality flag, '0' and '1' representing
the most probable star clusters. Some brighter star cluster candidates in the FSR catalogue have so far
been explored in detail. FSR\,1735 (\citealt{FSR1735}) and FSR\,1767 (\citealt{FSR1767}) are new globular
clusters in the Galaxy. Available evidence indicates that FSR\,584 is a also a new globular cluster
(\citealt{FSR584}). FSR\,190 (\citealt{FSR190}) is either a globular cluster or a very old open cluster.
FSR\,1744, FSR\,89 and FSR\,31 are old OCs in the inner Galaxy (\citealt{OldOCs}). Ruprecht\,101
(FSR\,1603) resulted to be an old OC, and FSR\,1755 an embedded cluster in the HII region Sh2-3
(\citealt{F1603}).

Another interesting approach is to explore complete samples of candidates within limited regions along
the disk. Recently, \citet{ProbFSR} carried out a systematic colour-magnitude (CMD) and structural
(with stellar radial density profiles - RDPs) analysis of FSR cluster candidates in bulge/disc
directions, at $|\ell|\le60^\circ$. With quality flags Q=0 and 1, the
complete sample contained 20 star cluster candidates. The results indicated 4 new and 2 previously known
OCs with ages in the
range 0.6\,Gyr to $\sim5$\,Gyr and distances from the Sun $\rm1.3\la\ds(kpc)\la2.8$, 5 uncertain cases
(that require deeper observations), and 9 probable field fluctuations.

In the present paper, 28 anticentre FSR cluster candidates are explored. Our approach is based
on 2MASS photometry, on which we apply a field-star decontamination algorithm (\citealt{BB07}) that is
essential to disentangle physical from field CMD sequences. We also take into account properties of the
stellar RDPs. A fundamental question to be addressed with this work is what fraction
of the candidates will turn out to be clusters, uncertain cases that require deeper observations, and
field fluctuations, as compared to a similar study of the disk in the opposite (bulge) direction 
(\citealt{ProbFSR}). In principle, more
clusters are expected owing to the lower level of crowding, field contamination, and absorption, and
because optical OCs appear to intrinsically populate more the anticentre regions (\citealt{DiskProp},
and references therein).

As will be seen in Sect.~\ref{targets}, several of the present sample candidates have a counterpart 
in optical studies, as e.g. in the WEBDA database. Recently, \citet{GKZ07} analysed a
$16^\circ\times16^\circ$ region around the anticentre and, likewise \citet{FSRcat}, they searched for 
new clusters or
candidates. They determined cluster astrophysical parameters for part of the sample. In the present study
we point out the clusters in common between \citet{FSRcat} and \citet{GKZ07}, which are part of the present
sample. Thus, we will have available astrophysical parameters of optical clusters to compare with presently
derived ones via 2MASS. It is very important to compare parameter determinations for clusters studied with
different datasets, or the same data with different techniques. The anticentre provides this opportunity.
We also point out that the availability of automated searches has provided elusive candidates that can
hardly or cannot be recognised at all by eye inspection like classical optical OCs, owing to field
contamination or absorption. Only decontamination methods can show their cluster nature or not (e.g.
\citealt{ProbFSR}).

This paper is structured as follows. In Sect.~\ref{targets} we provide fundamental data of the sample
candidates and present relevant data of the previously known OCs. In Sect.~\ref{PhotPar} we present the
2MASS photometry and discuss the methods employed in the CMD analyses, especially the field-star
decontamination. In Sect.~\ref{RDPs} we analyse the stellar radial density profiles and derive structural
parameters of the confirmed star clusters. In Sect.~\ref{Disc} we discuss the star cluster parameters as
a function of the position in the Galaxy. Concluding remarks are given in Sect.~\ref{Conclu}.

\section{The anti-centre FSR star cluster candidates}
\label{targets}

For this study we selected all cluster candidates with quality flags '0' and '1' projected within
$160^\circ\le\ell\le200^\circ$ and $-20^\circ\le b\le20^\circ$, taken from both classes of probable 
and possible candidates (\citealt{FSRcat}). Observational data on the sample targets are given in 
Table~\ref{tab1}. Also included are the core and tidal radii measured by \citet{FSRcat} on the 2MASS 
\hh\ images by means of a \citet{King1962} profile fit, and the quality flag. Table~\ref{tab1} 
separates the candidates according to the FSR classification as probable or possible star cluster.

\begin{table}
\caption[]{General data on the FSR star cluster candidates}
\label{tab1}
\renewcommand{\tabcolsep}{1.2mm}
\renewcommand{\arraystretch}{1.25}
\begin{tabular}{lccccccc}
\hline\hline
Target&$\alpha(2000)$&$\delta(2000)$&$\ell$&$b$&\rc&\rt&Q\\
&(hms)&($^\circ\,\arcmin\,\arcsec$)&($^\circ$)&($^\circ$)&(\arcmin)&(\arcmin)\\
(1)&(2)&(3)&(4)&(5)&(6)&(7)&(8)\\
\hline
\multicolumn{8}{c}{Probable Star Clusters}\\
\hline
FSR\,744 & 04:59:30 & $+$38:00:42 & 167.1 & $-2.1$ & 0.8 & 42.7 & 1 \\
FSR\,793 & 05:24:21 & $+$32:36:13 & 174.4 & $-1.8$ & 1.2 &  8.2 & 0 \\
FSR\,810 & 05:40:57 & $+$32:16:16 & 176.6 & $+0.9$ & 1.7 & 5.2 & 1 \\
FSR\,814 & 05:36:49 & $+$31:12:42 & 177.1 & $-0.4$ & 1.2 & 33.2 & 0 \\
FSR\,834 & 05:50:07 & $+$28:53:28 & 180.5 & $+0.8$ & 0.8 &  4.2 & 1 \\
FSR\,855 & 05:42:22 & $+$22:49:34 & 184.8 & $-3.8$ & 0.9 & 11.1 & 0 \\
FSR\,869 & 06:10:05 & $+$24:33:31 & 186.6 & $+2.5$ & 0.8 & 11.7 & 1 \\
FSR\,894 & 06:04:05 & $+$20:16:51 & 189.6 & $-0.7$ & 1.0 & 25.3 & 0 \\
FSR\,911 & 06:25:00 & $+$19:52:03 & 192.3 & $+3.4$ & 0.8 &  7.1 & 0 \\
FSR\,923 & 06:10:36 & $+$16:58:16 & 193.2 & $-1.0$ & 0.8 &  7.3 & 1 \\
FSR\,927 & 06:24:10 & $+$18:01:30 & 193.8 & $+2.3$ & 1.2 & 16.4 & 1 \\
FSR\,932 & 06:04:24 & $+$14:33:43 & 194.6 & $-3.5$ & 0.7 & 25.9 & 1 \\
FSR\,942 & 06:05:58 & $+$13:40:06 & 195.6 & $-3.6$ & 0.9 & 43.9 & 1 \\
FSR\,948 & 06:25:59 & $+$15:51:08 & 196.0 & $+1.7$ & 0.8 & 18.7 & 1 \\
FSR\,956 & 06:12:25 & $+$13:00:26 & 196.9 & $-2.5$ & 0.8 & 5.9 & 1 \\
FSR\,974 & 06:32:38 & $+$12:33:19 & 199.6 & $+1.6$ & 1.1 &  7.8 & 1 \\
\hline
\multicolumn{8}{c}{Possible Star Clusters}\\
\hline
FSR\,705 & 05:11:43 & $+$47:41:42 & 160.7 & $+4.9$ & 1.3 & 8.0 & 0 \\
FSR\,729 & 05:25:55 & $+$46:29:46 & 163.1 & $+6.2$ & 1.0 & 5.2 & 1 \\
FSR\,730 & 06:02:33 & $+$49:52:24 & 163.2 & $+13.1$& 0.8 & 37.0 & 1 \\
FSR\,756 & 04:24:16 & $+$29:43:31 & 168.6 & $-13.7$& 0.8 & 17.0 & 1 \\
FSR\,773 & 04:29:40 & $+$26:01:04 & 172.3 & $-15.3$& 1.0 & 6.2 & 1 \\
FSR\,776 & 06:07:28 & $+$39:50:23 & 172.7 & $+9.3$ & 1.3 & 7.9 & 1 \\
FSR\,801 & 04:46:57 & $+$24:53:16 & 175.8 & $-13.0$& 1.9 & 7.6 & 1 \\
FSR\,841 & 05:06:18 & $+$21:31:00 & 181.3 & $-11.5$& 0.8 & 43.5 & 1 \\
FSR\,851 & 05:14:39 & $+$19:48:01 & 183.8 & $-10.9$& 1.2 & 31.7 & 0 \\
FSR\,882 & 05:27:54 & $+$16:54:32 & 188.1 & $ -9.8$& 0.7 &  9.5 & 0 \\
FSR\,884 & 05:32:21 & $+$17:11:02 & 188.4 & $ -8.8$& 0.8 & 5.2 & 1 \\
FSR\,917 & 06:33:17 & $+$20:31:08 & 192.6 & $ +5.4$& 0.8 & 38.5 & 1 \\
\hline
\end{tabular}
\begin{list}{Table Notes.}
\item Cols.~2-3: Central coordinates provided by \citet{FSRcat}. Cols.~4-5: Corresponding Galactic
coordinates. Cols.~6 and 7: Core and tidal radii derived by \citet{FSRcat} from King fits to the
2MASS \hh\ images. Col.~8: FSR quality flag.
\end{list}
\end{table}

Seven open clusters in the present FSR subsample have previous literature identifications (WEBDA), while
2 have been subsequently identified by \citet{GKZ07}. Table~\ref{tab2} shows the cross-identifications 
and relevant references. We will adopt the first designation throughout this paper. Finally, we note that
FSR\,911 is located $\approx8\arcmin$ to the northwest of the young stellar system Bochum\,1 (\citealt{Moffat75};
\citealt{Yadav03}). Indeed, a deeper analysis of the region of FSR\,911 suggested that it is not Bochum\,1,
as discussed in \citet{Boch1}.

\begin{table}
\caption[]{Cross-identification of the open clusters}
\label{tab2}
\renewcommand{\tabcolsep}{1.4mm}
\renewcommand{\arraystretch}{1.25}
\begin{tabular}{lllll}
\hline\hline
Desig\#1&Desig\#2&Desig\#3&Desig\#4&References\\
(1)&(2)&(3)&(4)&(5)\\
\hline
Berkeley\,23 & FSR\,917      &   ---         &  ---     & 1,2\\
Berkeley\,69 & FSR\,793      &   ---         &  ---     & 1,2\\
Berkeley\,71 & FSR\,810      &   ---         &  ---     & 1,2\\
Czernik\,23  & FSR\,834      &   ---         &  ---     & 1,2\\
NGC\,1798    & Berkeley\,16  & FSR\,705      &  ---     & 1,1,2\\
NGC\,1883    & Collinder\,64 & FSR\,729      &  ---     & 1,1,2\\
NGC\,2126    & Melotte\,39   & Collinder\,78 & FSR\,730 & 1,1,1,2\\
FSR\,814     & Koposov\,36   &   ---         &  ---     & 2,3\\
FSR\,869     & Koposov\,63   &   ---         &  ---     & 2,3\\
 \hline
\end{tabular}
\begin{list}{Table Notes.}
\item (1) - \citet{Alter70}; (2) - \citet{FSRcat}; (3) - \citet{GKZ07}.
\end{list}
\end{table}
 
Previous studies with results relevant to the present paper are summarised below.

{\tt NGC\,1798:} This object is an IAC. \citet{Park99} derived the age $\tau=1.4\pm0.2$\,Gyr,
distance from the Sun $\ds=4.2\pm0.3$\,kpc, reddening $\ebv=0.51\pm0.04$, and limiting
radius $R_{lim}=8\farcm3\approx10.2$\,pc. \citet{Lata02} found $\tau=1.4$\,Gyr and $\ebv=0.51$.
\citet{MacNie07} found $\tau=1.6$\,Gyr, $\ebv=0.37\pm0.10$, $\ds=3.55\pm0.7$\,kpc,
$R_{lim}=9\arcmin$, core radius $\rc=1\farcm3\pm0\farcm1$, and mass $M=6932\,\ms$.

{\tt NGC\,1883:} \citet{Tad2002} found $R_{lim}=6\arcmin$, $\rc=0\farcm33$, and $M=480 - 650\,\ms$.
\citet{Carraro03} derived $\tau\sim1$\,Gyr, $\ebv=0.23 - 0.35$, $\ds=4.8$\,kpc, and $R_{lim}=2\farcm5$.
\citet{Villa07} derived the age $\tau=650\pm70$\,Myr.

{\tt NGC\,2126:} \citet{Gaspar03} derived $\tau\sim1$\,Gyr, $\ebv=0.20\pm0.15$, and 
$\ds=1.3\pm0.6$\,kpc. \citet{MacNie07} found $\tau=1.3$\,Gyr, $\ebv=0.27\pm0.11$, $\ds=1.1\pm0.4$\,kpc,
$R_{lim}=10\arcmin$, $\rc=1\farcm9\pm0\farcm3$, and $M=395\,\ms$.

{\tt Berkeley\,69:} \citet{Durga01} found $\rc=49\farcs3\pm3\farcs9$, $R_{lim}=110\arcsec$, and
$M=84\,\ms$.

{\tt Berkeley\,71:} \citet{Lata04} found $\tau=320$\,Myr, $\rc=1\farcm6\pm0\farcm3$, and
$R_{lim}=3.4$\,pc. \citet{MacNie07} derived $\tau=1$\,Gyr, $\ebv=0.81\pm0.08$, $\ds=3.26\pm0.7$\,kpc,
$R_{lim}=3\farcm3=3.1\pm0.7$\,pc, $\rc=1\farcm2\pm0\farcm2=1.11\pm0.44$\,pc, and $M=256\,\ms$.

{\tt FSR\,814:} \citet{GKZ07} derived $\tau<31$\,Myr, $\ebv=0.91\pm0.16$, and $\ds=1.5\pm0.1$\,kpc.

{\tt Czernik\,23:} \citet{GKZ07} derived $\tau=282\pm50$\,Myr, $\ebv=0.38\pm0.02$, and $\ds=2.5\pm0.1$\,kpc.

{\tt FSR\,869:} \citet{GKZ07} derived $\tau=1.4\pm0.1$\,Gyr, $\ebv=0.26\pm0.04$, and $\ds=3.0\pm0.3$\,kpc.

{\tt Berkeley\,23:} \citet{Ann02} found $\tau=794$\,Myr, $\ebv=0.40$, and $\ds=6.9$\,kpc.
\citet{Hasegawa04} derived $\tau=1.8$\,Gyr.

Recently, Cz\,23 has been studied in detail by \citet{Cz23}, who have derived the parameters
$\tau=4.5\pm0.5$\,Gyr, $\ebv=0.0\pm0.1$, $\ds=2.5\pm0.1$\,kpc, $\rc=0\farcm49\pm0\farcm04=0.36\pm0.08$\,pc,
$R_{lim}=4\farcm9\pm0\farcm7=3.6\pm0.4$\,pc, and a total mass of $M\sim115\,\ms$.

\begin{figure}
\resizebox{\hsize}{!}{\includegraphics{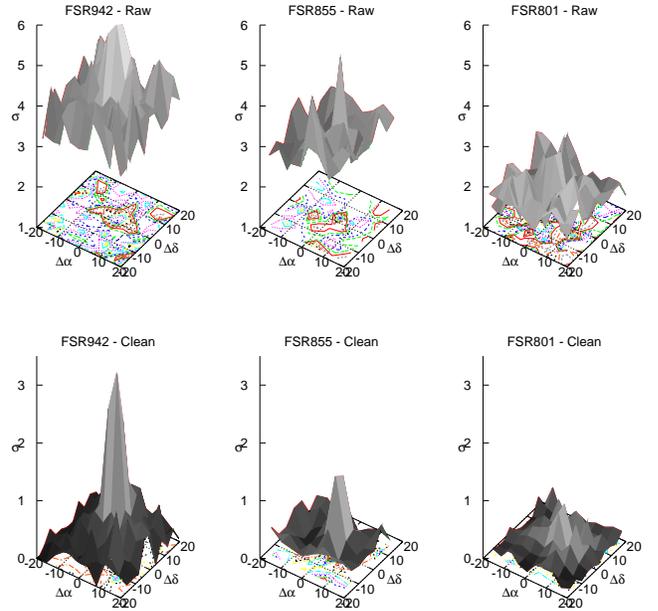}}
\caption[]{Stellar surface-density ($\sigma$, in units of $\rm stars\ arcmin^{-2}$) of representative 
examples of an OC identified in this work (FSR\,942, left panels), uncertain case (FSR\,855, middle), 
and possible field fluctuation (FSR\,801, right). The surfaces were built for a mesh size of 
$4\arcmin\times 4\arcmin$, centred on the coordinates in Table~\ref{tab1}. The observed (raw) and 
field-star decontaminated photometry are shown in the top and bottom panels, respectively.}
\label{fig1}
\end{figure}

\begin{figure}
\resizebox{\hsize}{!}{\includegraphics{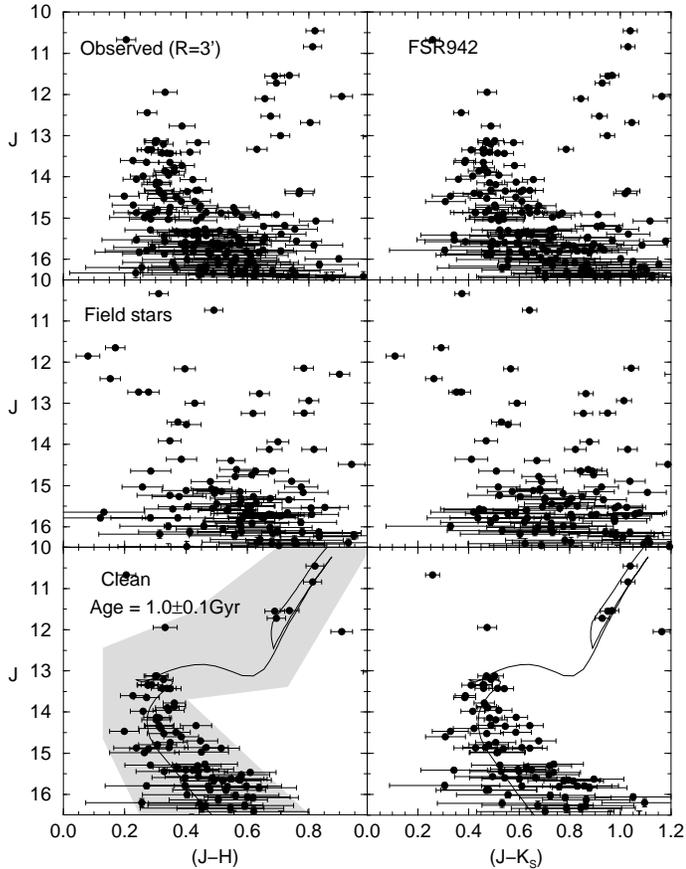}}
\caption[]{2MASS CMDs extracted from the $R<3\arcmin$ region of the OC FSR\,942. Top panels: observed photometry
with the colours $\jj\times\jh$ (left) and $\jj\times\jk$ (right). A relatively populous MS and a giant
clump are suggested, together with an important contamination by disk  stars. Middle: equal-area comparison
field extracted from the region $19\farcm77 - 20\arcmin$.  Bottom panels: decontaminated CMDs with the
1.0\,Gyr Padova isochrone (solid line), showing the enhanced cluster morphology. The colour-magnitude filter
used to isolate cluster MS/evolved stars is shown as a shaded region.}
\label{fig2}
\end{figure}

\section{Photometry and analytical tools}
\label{PhotPar}

In this section we briefly describe the photometry and outline the methods we apply in the CMD
analyses.

\subsection{2MASS photometry}
\label{2mass}

2MASS photometry in the \jj, \hh\ and \ks\ bands was extracted in all cases in a relatively wide circular 
field of 30\arcmin\ in radius, centred on the coordinates provided by \citet{FSRcat} (cols.~2
and 3 of Table~\ref{tab1}). Photometry extraction was performed with the VizieR\footnote{\em vizier.u-strasbg.fr/viz-bin/VizieR?-source=II/246} tool.
Wide extraction areas are necessary for statistical representativity of magnitude and colours, for a
consistent field star decontamination (Sect.~\ref{FSD}). They are important as well for stellar radial
density profiles with a high contrast with respect to the background (Sect.~\ref{RDPs}). Properties and
limitations of wide 2MASS-extraction areas are discussed in detail in, e.g. \citet{BB07}. In some cases
the RDP built with the original FSR coordinates presented a dip at the centre. Consequently, new central
coordinates were searched to maximise the star-counts in the innermost RDP bin. The optimised central 
coordinates are given in cols.~2 and 3 of Table~\ref{tab5}.

As a photometric quality constraint, the 2MASS extractions were restricted to stars with errors in \jj,
\hh\ and \ks\ smaller than 0.25\,mag. About $75\% - 85\%$ of the stars in all extractions considered here
have errors smaller than 0.06\,mag. A typical distribution of 2MASS uncertainties as a function of magnitude,
for objects projected towards the central parts of the Galaxy, can be found in \citet{BB07}. 

\begin{table*}
\caption[]{Statistics of the field-star decontamination, in magnitude bins, for representative cases}
\label{tab3}
\renewcommand{\tabcolsep}{1.02mm}
\renewcommand{\arraystretch}{1.25}
\begin{tabular}{ccccccccccccccccccc}
\hline\hline
&\multicolumn{5}{c}{FSR\,942 ($R<3\arcmin$)}&&\multicolumn{5}{c}{FSR\,855 ($R<3\arcmin$)}
&&\multicolumn{5}{c}{FSR\,801 ($R<3\arcmin$)}\\
\cline{2-6}\cline{8-12}\cline{14-18}
$\Delta\jj$&\no&\nc&\ns&\sFS&\fsU && \no&\nc&\ns&\sFS&\fsU&& \no&\nc&\ns&\sFS&\fsU\\
  (mag)    &(stars)&(stars)&&(stars)& && (stars)&(stars)&&(stars)& && (stars)&(stars)&&(stars)\\
\cline{1-18}
  7--8& --- & --- & --- & --- & ---&&$1\pm1.0$ &1  &1.0&0.42&1.07&&---&---&---&---&---\\
  8--9& --- & --- & --- & --- & ---&& ---      &---&---&--- &--- &&---&---&---&---&---  \\
 9--10& --- & --- & --- & --- & ---&&$1\pm1.0$ &1  &1.0&0.99&1.42&&---&---&---&---&---    \\
10--11&$3\pm1.7$ &3 &1.7&0.57&0.51 &&$5\pm2.2$ &1  &0.4&1.12&1.02&&---&---&---&---&--- \\
11--12&$4\pm2.0$ &4 &2.0&0.38&0.18 &&$4\pm2.0$ &2  &1.0&1.70&0.49&&$2\pm1.4$&1 &0.7&0.18&0.23 \\
12--13&$7\pm2.6$ &1 &0.4&0.50&0.12 &&$6\pm2.4$ &1  &0.4&3.09&0.43&&$4\pm2.0$&2 &1.0&0.30&0.20  \\
13--14&$24\pm4.9$&17&3.5&0.75&0.08 &&$20\pm4.5$&3  &0.7&2.47&0.19&&$4\pm2.0$&2 & 1.0&0.36&0.10 \\
14--15&$38\pm6.2$&21&3.4&1.61&0.09 &&$33\pm5.7$&6  &1.0&6.27&0.26&&$9\pm3.0$&3&1.0&0.72&0.11  \\
15--16&$63\pm7.9$&27&3.4&2.87&0.08 &&$56\pm7.5$&12 &1.6&5.31&0.11&&$25\pm5.0$&9&1.8&1.60&0.12  \\
16--17&$45\pm6.7$&21&3.1&4.91&0.14 &&$72\pm8.5$&29 &3.4&6.14&0.12&&$21\pm4.6$&8&1.7&1.78&0.09 \\
\cline{2-6}\cline{8-12}\cline{14-18}
All &$184\pm13.6$&94&6.8&3.2&0.03& &$198\pm14.1$&56&3.6&11.1&0.07 &&$65\pm8.1$&25&2.9&2.8&0.06\\
\hline
\end{tabular}
\begin{list}{Table Notes.}
\item The table provides, for each magnitude bin ($\Delta\jj$), the number of observed stars
(\no) within the spatial region sampled in the CMDs shown in the top panels of Figs.~\ref{fig3}
and \ref{fig4}, the respective number of probable member stars (\nc) computed by the decontamination 
algorithm, the \ns\ parameter, the $\rm 1\,\sigma$ Poisson fluctuation (\sFS) around the mean, with 
respect to the star counts measured in the 8 sectors of the comparison field, and the field-star
uniformity parameter (\fsU). The statistical significance of \nc\ is reflected in its ratio with
the $1\sigma$ Poisson fluctuation of \no\ (\ns) and with \sFS. The bottom line corresponds to the
full magnitude range.
\end{list}
\end{table*}

\begin{figure}
\resizebox{\hsize}{!}{\includegraphics{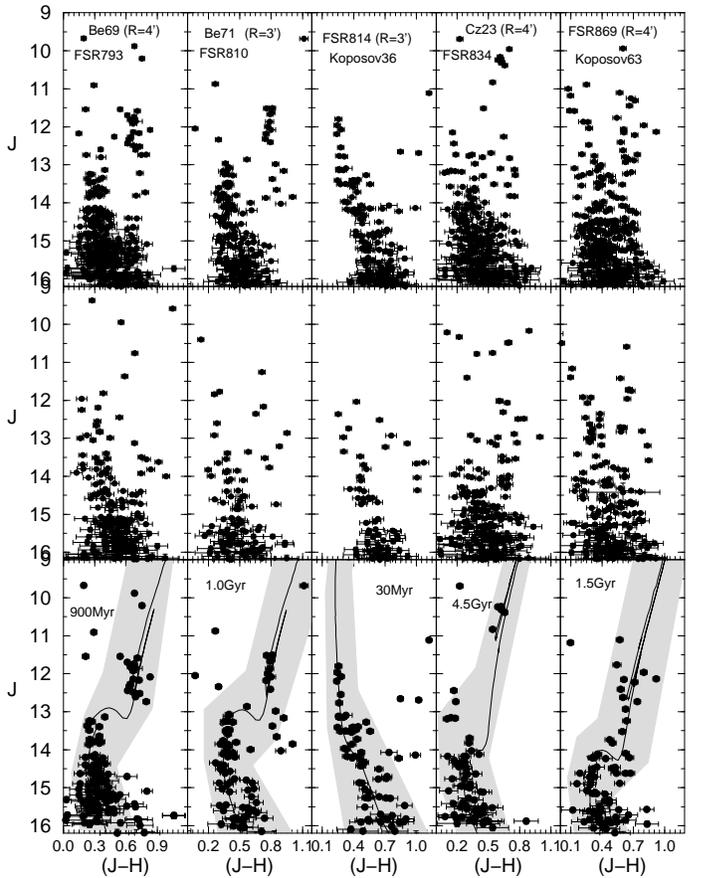}}
\caption{Same as Fig.~\ref{fig2} for the first half of the subsample of previously known
OCs included in this paper. Only the $\jj\times\jh$ CMD is shown.}
\label{fig3}
\end{figure}

\begin{figure}
\resizebox{\hsize}{!}{\includegraphics{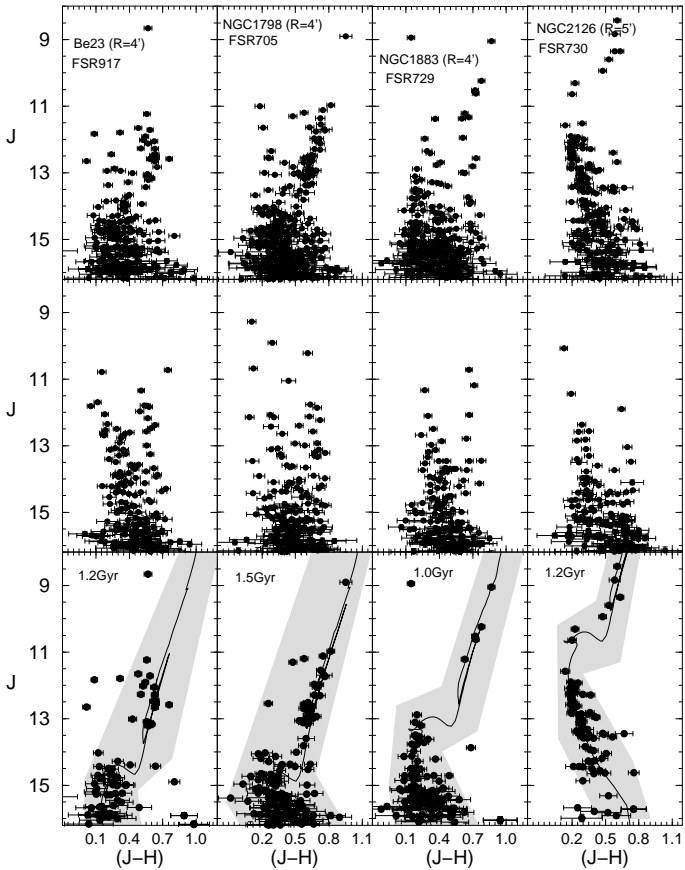}}
\caption{Same as Fig.~\ref{fig3} for the remaining previously known OCs.}
\label{fig4}
\end{figure}

\begin{figure}
\resizebox{\hsize}{!}{\includegraphics{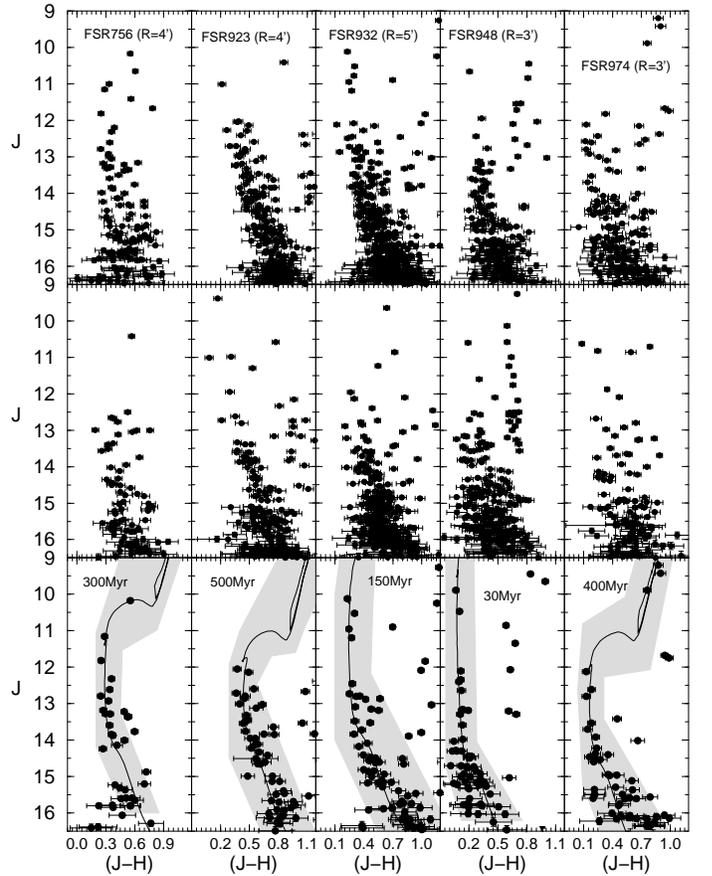}}
\caption{Same as Fig.~\ref{fig3} for the remaining OCs identified here. FSR\,942 is shown
in Fig.~\ref{fig2}.}
\label{fig5}
\end{figure}

\begin{figure}
\resizebox{\hsize}{!}{\includegraphics{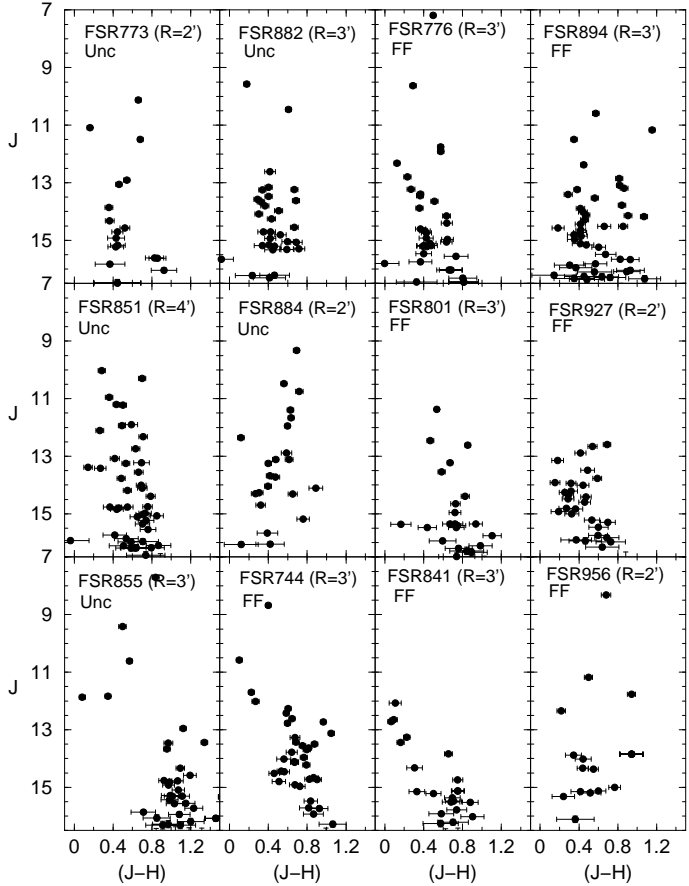}}
\caption{Field-star decontaminated $\jj\times\jh$ CMDs of the uncertain cases (Unc) and the 
possible field fluctuations (FF).}
\label{fig6}
\end{figure}

\begin{table}
\caption[]{Field-star decontamination: integrated statistics}
\label{tab4}
\renewcommand{\tabcolsep}{1.5mm}
\renewcommand{\arraystretch}{1.2}
\begin{tabular}{lcccccc}
\hline\hline
Object&${\rm R_{max}}$&\no&\nc&\ns&\sFS&\fsU \\
  &(\arcmin)&(stars)&(stars)&&(stars) \\
\hline
&\multicolumn{6}{c}{Confirmed OCs}\\
\cline{2-7}
Be\,23&4&$222\pm14.9$&67&4.5&10.9&0.06 \\
Be\,69  &4&$308\pm17.5$&134&7.6& 9.6&0.05 \\
Be\,71  &3&$216\pm14.7$&104&7.1& 6.0&0.05 \\
Cz\,23&4&$228\pm14.1$& 66&4.4&15.6&0.07 \\
NGC\,1798&4&$372\pm19.3$&156&8.1&13.3&0.06 \\
NGC\,1883&4&$291\pm17.1$&115&6.7&7.5&0.04 \\
NGC\,2126&5&$227\pm15.1$& 76&5.0&6.8&0.04 \\
FSR\,756&4&$141\pm11.9$&44&3.7&5.2&0.05 \\
FSR\,814&3&$187\pm13.7$& 89&6.5& 5.4&0.04 \\
FSR\,869&4&$378\pm19.4$& 96&4.9&23.1&0.08 \\
FSR\,923&4&$256\pm16.0$&69&4.3&14.1&0.06 \\
FSR\,932&5&$361\pm19.0$&81&4.2&7.1&0.02 \\
FSR\,942&3&$184\pm13.6$&94&6.8&3.2&0.03 \\
FSR\,948&4&$280\pm16.7$&66&3.9&9.3&0.04 \\
FSR\,974&3&$180\pm13.4$&67&5.0&6.5&0.05 \\
\cline{2-7}
&\multicolumn{6}{c}{Uncertain cases}\\
\cline{2-7}
FSR\,773&2&$33\pm5.7$&16&2.8&2.0&0.09 \\
FSR\,851&4&$158\pm12.6$&49&3.9&7.5&0.06 \\
FSR\,855&3&$198\pm14.1$&56&3.6&11.1&0.07 \\
FSR\,882&3&$93\pm9.6$&31&3.2&2.6&0.04 \\
FSR\,884&2&$47\pm6.8$&18&2.7&1.9&0.06 \\
FSR\,911&4&$310\pm17.6$& 59&3.3&17.8&0.07 \\
\cline{2-7}
&\multicolumn{6}{c}{Possible field fluctuations}\\
\cline{2-7}
FSR\,744&3&$144\pm12.0$&37&3.1&3.3&0.03 \\
FSR\,776&3&$105\pm10.2$&39&3.8&2.7&0.04 \\
FSR\,801&3&$65\pm8.1$&25&2.9&2.8&0.06\\
FSR\,841&3&$72\pm8.5$&25&2.9&3.5&0.07 \\
FSR\,894&3&$169\pm13.0$&42&3.2&8.0&0.06 \\
FSR\,927&2&$84\pm9.2$&32&3.5&4.8&0.08 \\
FSR\,956&2&$65\pm8.1$&13&1.6&2.0&0.04 \\
\hline
\end{tabular}
\begin{list}{Table Notes.}
\item Same as Table~\ref{tab5} for the statistics of the full magnitude range covered
by the respective CMD. CMDs extracted from $\rm 0\la R(\arcmin)\la R_{max}$.
\end{list}
\end{table}

In Fig.~\ref{fig1} we show the spatial distribution of the stellar surface-density, as measured with the
2MASS photometry, for representative cases of the three different types of objects dealt with in this work.
Surface-densities computed with the observed (top panels) and field-star decontaminated (Sect.~\ref{FSD})
photometries (bottom) are included. A confirmed open cluster (FSR\,942), an uncertain case (FSR\,855), and a
possible field fluctuation (FSR\,801) are illustrated in Fig.~\ref{fig1}. This figure shows the surface density
($\sigma$, in units of $\rm stars\,arcmin^{-2}$) for a rectangular mesh with cells of dimensions
$4\arcmin\times4\arcmin$. The mesh extends up to $|\Delta\alpha|=|\Delta\delta|\approx20\arcmin$ with respect
to the centre, in right ascension and declination. Most of the cluster structure is contained in the central
cell, especially for FSR\,942 and FSR\,855. The rather irregular surface-density distribution produced with
the observed photometry, especially for FSR\,801, occurs because of the important contamination by disk stars.
Even so, a central excess can be seen, which corresponds to the overdensity detected by \citet{FSRcat}. FSR\,942,
on the other hand, clearly detaches in the central cells (top-left panel) against a less irregular field.

\subsection{Colour-magnitude diagrams}
\label{CMDs}

CMDs are fundamental for establishing the nature of the candidates in the present analysis. To illustrate this 
procedure we show in Fig.~\ref{fig2} the 2MASS $\jj\times\jh$ and $\jj\times\jk$ CMDs extracted from a
central ($R<3\arcmin$) region of FSR\,942. This CMD, which corresponds to $\approx30\%$ of the radial
density profile radius \rl\ (Sect.~\ref{RDPs}), shows a cluster-like population (the MS and a giant clump)
mixed with a component of disk stars. A first assessment on the relative fraction of the contamination is
provided by the equal-area comparison field (middle panels), extracted from the ring located at $19\farcm77
- 20\arcmin$. Differences in densities between the observed and comparison field CMDs  further suggest the
presence of a main sequence and a giant clump of an intermediate-age OC. Similar features are present in the
$\jj\times\jk$ CMD (top-right panel).

The observed CMDs of the remaining objects, extracted from central regions, are shown in the top panels 
of Figs.~\ref{fig3} to \ref{fig5}, with the corresponding comparison-field CMDs at the middle panels. For
the sake of space, only the \jh\ CMDs are shown. Similarly to the case of FSR\,942, essentially the same
CMD features are present in both colours.

\subsection{Field-star decontamination}
\label{FSD}

Field stars, mostly from the disk, contribute in varying proportions to the CMDs of the present 
objects (Figs.~\ref{fig2} - \ref{fig5}). In some cases, it appears to be the
dominant component. Thus, it is essential to quantify the relative densities of field stars and
potential cluster sequences to settle the nature of the cluster candidates, whether they are 
physical systems or field fluctuations.

We tackle this issue with the statistical algorithm described in detail in \cite{BB07} and
\citet{ProbFSR}. We present here only a brief description of the algorithm. It computes the
relative number-densities of probable field and cluster stars in cubic cells with axes along 
the \jj\ magnitude and the \jh\ and \jk\ colours. Stars are subtracted from each cell in a 
number that corresponds to the number-density of field stars measured within the same cell
in the comparison field. Typical cell dimensions are $\Delta\jj=0.5$, and $\Delta\jh=\Delta\jk=0.25$, 
which are a compromise between cell width and CMD resolution. They are wide enough to allow sufficient 
star-count statistics in individual cells and small enough to preserve the morphology of different
CMD sequences. A wide ring beyond the RDP
radius (Sect.~\ref{RDPs}) is used as comparison field, to provide a representative field star-count
statistics. Note that the equal-area extractions shown in the middle panels of Figs.~\ref{fig2} to
\ref{fig5} serve only for visual comparisons between central and offset field CMDs. The actual
decontamination process is carried out with the wide surrounding ring as described above. Further 
details on the algorithm, including discussions on subtraction efficiency and limitations, are given
in \citet{BB07}.

As output of the algorithm we have \nc\, the number of probable (i.e. decontaminated) member
stars, and the parameter \ns\ which, for a given spatial extraction, corresponds to the ratio of
\nc\ with respect to the corresponding $\rm1\sigma$ Poisson fluctuation of the number of observed
stars. For instance, the number (and uncertainty) of observed stars in the $R<3\arcmin$ CMD of FSR\,942
(Fig.~\ref{fig2} and Table~\ref{tab4}) is $\rm N_{obs}\pm\sigma_{N_{obs}}=184\pm13.6$, while the
corresponding number of decontaminated stars is $\nc=94$. Thus, we derive for this extraction
$\ns=\nc/\sigma_{N_{obs}}=6.8$. In this sense, \ns\ gives a measure of the statistical significance of
the decontaminated number of stars. Both parameters, \nc\ and \ns, can be computed for the full
range of magnitude covered by the CMD, or in individual magnitude ranges (see below). CMDs of star
clusters have \ns\ significantly higher than 1 (\citealt{ProbFSR}).
The algorithm also computes \sFS, which corresponds to the $\rm 1\,\sigma$ Poisson fluctuation
around the mean of the star counts measured in the 8 equal-area sectors of the comparison field. Thus, 
\sFS\ measures the spatial uniformity of the star counts in the comparison field. Low values of \sFS\
are expected in a uniform comparison field. Ideally, star clusters should have \nc\ higher than 
$\sim3\,\sFS$. It also computes \fsU, which measures the star-count uniformity of the comparison 
field, defined as $\rm\fsU=\sigma_{\langle N\rangle}/\langle N\rangle$, where $\langle N\rangle$ 
and $\sigma_{\langle N\rangle}$ are the average and standard deviation of the number of stars over 
all sectors. Non uniformities such as heavy differential reddening should result in high values of
\fsU.

Since we usually work with larger comparison fields than the cluster extractions, the correction
for the different spatial areas between field and cluster is expected to produce a fractional number
of probable field stars ($n_{fs}^{cell}$) in some cells. Before the cell-by-cell subtraction, the 
fractional numbers are rounded off to the nearest integer, but limited to the number of observed stars 
in each cell ($n_{sub}^{cell}=NI(n_{fs}^{cell})\leq n_{obs}^{cell}$, where NI represents rounding off to 
the nearest integer). The global effect is quantified by means of the difference between the expected number 
of field stars in each cell ($n_{fs}^{cell}$) and the actual number of subtracted stars ($n_{sub}^{cell}$). 
Summed over all cells, this quantity provides an estimate of the total subtraction efficiency of the 
process, \[ f_{sub}=100\times\sum_{cell}n_{sub}^{cell}/\sum_{cell}n_{fs}^{cell}~~(\%).\] Ideally, the best 
results would be obtained for an efficiency $f_{sub}\approx100\%$. The adopted grid settings produced 
subtraction efficiencies higher than 93\% in all cases. 

Table~\ref{tab3} presents the full statistics of the decontamination of FSR\,942,
FSR\,855, and FSR\,801. The parameters discussed above are presented in magnitude bins and for the
full CMD magnitude range, which allows verification of dependences with magnitude. For the remaining
cases only the integrated statistics is given in Table~\ref{tab4}. The decontaminated CMDs are shown
in the bottom panels of Figs.~\ref{fig2} - \ref{fig5}, and in Fig.~\ref{fig6}.

Based on the decontaminated CMDs and the RDP properties (Sect.~\ref{RDPs}), the objects can be grouped
into three different classes, {\em (i)} confirmed OCs (with 15 objects), {\em (ii)} uncertain cases (6), 
and {\em (iii)} possible field fluctuations (7). Table~\ref{tab4} is arranged according to these classes. 
As expected, the integrated \ns\ parameter of the confirmed OCs is higher than 3.0, in some cases reaching 
$\approx8$. The uncertain cases and the possible field fluctuations have lower values, around 3.

Besides the 7 previously known and the 2 recently identified OCs (Table~\ref{tab2}), 6 new ones show up 
in this work: FSR\,756, FSR\,923, FSR\,932, FSR\,942, FSR\,948, and FSR\,974. 

\subsection{Fundamental parameters}
\label{FundPar}

Fundamental parameters are derived for the cases where a significant probability of a star cluster 
occurs. To this purpose we fit the decontaminated CMDs with solar-metallicity Padova isochrones
(\citealt{Girardi02}) computed with the 2MASS \jj, \hh\ and \ks\ filters\footnote{\em
stev.oapd.inaf.it/$\sim$lgirardi/cgi-bin/cmd }. The 2MASS transmission filters produced isochrones 
very similar to the Johnson-Kron-Cousins (e.g. \citealt{BesBret88}) ones, with differences of at most
0.01 in \jh\ (\citealt{TheoretIsoc}).

The isochrone fit gives the age and the reddening \ejh, which converts to \ebv\ and \aV\ through the
transformations $A_J/A_V=0.276$, $A_H/A_V=0.176$, $A_{K_S}/A_V=0.118$, and $A_J=2.76\times\ejh$
(\citealt{DSB2002}), for a constant total-to-selective absorption ratio $R_V=3.1$. These ratios
were derived from the extinction curve of \citet{Cardelli89}. We also compute
the distance from the Sun (\ds) and the Galactocentric distance (\dgc), based on the recently derived value
of the Sun's distance to the Galactic centre $\Rgc=7.2$\,kpc derived with updated parameters of Globular
clusters (\citealt{GCProp}). Age, \aV, \ds\ and \dgc\ are
given in cols.~4 to 7 of Table~\ref{tab5}, respectively. The isochrone fits to the probable star clusters
are shown in the bottom panels of Figs.~\ref{fig2} to \ref{fig5}.

\begin{table*}
\caption[]{Fundamental parameters derived in this work}
\label{tab5}
\renewcommand{\tabcolsep}{1.6mm}
\renewcommand{\arraystretch}{1.25}
\begin{tabular}{lccccccccc}
\hline\hline
Target&$\alpha(2000)$&$\delta(2000)$&Age&\aV&\ds&\dgc&\xgc&\ygc&\zgc\\
&(hms)&($^\circ\,\arcmin\,\arcsec$)&(Myr)&(mag)&(kpc)&(kpc)&(kpc)&(kpc)&(kpc)\\
(1)&(2)&(3)&(4)&(5)&(6)&(7)&(8)&(9)&(10)\\
\hline
\multicolumn{10}{c}{Confirmed star clusters}\\
\hline
Be\,23&06:33:16.2&$+$20:31:08.0&$1200\pm200$&$0.3\pm0.2$&$6.7\pm0.4$&$13.8\pm0.4$&$-13.7\pm0.4$&$-1.45\pm0.08$&$+0.62\pm0.03$  \\
Be\,69  &05:24:21.6&$+$32:36:3.2&$900\pm100$&$1.5\pm0.2$&$3.5\pm0.2$&$10.7\pm0.2$&$-10.7\pm0.2$&$+0.33\pm0.02$&$-0.11\pm0.01$  \\
Be\,71  &($\dagger$)&($\dagger$)&$1000\pm100$&$2.3\pm0.2$&$3.0\pm0.2$&$10.2\pm0.2$&$-10.2\pm0.2$&$+0.28\pm0.01$&$+0.05\pm0.01$ \\
Cz\,23&($\dagger$)&($\dagger$)&$4500\pm500$&$0.0\pm0.1$&$2.5\pm0.1$&$9.7\pm0.1$&$-9.7\pm0.1$&$-0.02\pm0.01$&$+0.04\pm0.01$  \\
NGC\,1798&($\dagger$)&($\dagger$)&$1500\pm300$&$0.8\pm0.2$&$6.0\pm0.3$&$13.0\pm0.3$&$-12.8\pm0.3$&$+1.97\pm0.10$&$+0.51\pm0.03$  \\
NGC\,1883&($\dagger$)&($\dagger$)&$1000\pm100$&$0.5\pm0.2$&$3.7\pm0.2$&$10.8\pm0.2$&$-10.8\pm0.2$&$+1.08\pm0.06$&$+0.40\pm0.02$  \\
NGC\,2126&06:02:34.6&$+$49:51:36.0&$1200\pm200$&$0.5\pm0.1$&$1.0\pm0.1$&$8.2\pm0.1$&$-8.2\pm0.1$&$+0.29\pm0.02$&$+0.23\pm0.01$  \\
FSR\,756&04:24:13.4&$+$29:42:14.4&$300\pm50$&$3.0\pm0.1$&$1.8\pm0.1$&$8.9\pm0.1$&$-8.9\pm0.1$&$+0.34\pm0.02$&$-0.43\pm0.02$  \\
FSR\,814&05:36:46.1&$+$31:11:45.6&$30\pm20$&$3.0\pm0.2$&$1.6\pm0.1$&$8.9\pm0.1$&$-8.9\pm0.1$&$+0.08\pm0.01$&$-0.01\pm0.01$ \\
FSR\,869&06:10:01.9&$+$24:32:54.6&$1500\pm300$&$1.3\pm0.1$&$4.2\pm0.2$&$11.4\pm0.2$&$-11.4\pm0.2$&$-0.48\pm0.02$&$+0.19\pm0.01$  \\
FSR\,923&($\dagger$)&($\dagger$)&$500\pm100$&$4.2\pm0.1$&$1.5\pm0.1$&$8.7\pm0.1$&$-8.7\pm0.1$&$-0.36\pm0.02$&$-0.03\pm0.01$  \\
FSR\,932&06:04:26.4&$+$14:33:20.2&$150\pm50$&$2.8\pm0.1$&$1.5\pm0.1$&$8.7\pm0.1$&$-8.7\pm0.1$&$-0.39\pm0.02$&$-0.09\pm0.01$  \\
FSR\,942&($\dagger$)&($\dagger$)&$1000\pm100$&$1.6\pm0.1$&$3.1\pm0.2$&$10.2\pm0.1$&$-10.2\pm0.1$&$-0.83\pm0.04$&$-0.19\pm0.01$  \\
FSR\,948&06:25:52.8&$+$15:50:15.0&$30\pm10$&$1.5\pm0.1$&$2.9\pm0.1$&$10.0\pm0.1$&$-10.0\pm0.1$&$-0.79\pm0.04$&$+0.08\pm0.01$  \\
FSR\,974&06:32:41.3&$+$12:31:55.2&$400\pm100$&$1.6\pm0.1$&$2.6\pm0.1$&$9.7\pm0.1$&$-9.7\pm0.1$&$-0.87\pm0.04$&$+0.07\pm0.01$  \\
\hline
\multicolumn{10}{c}{Uncertain cases: deeper photometry necessary}\\
\hline
FSR\,773&04:29:37.0&$+$26:00:14.0  \\
FSR\,851&05:14:44.9&$+$19:47:31.2  \\
FSR\,855&05:42:21.6&$+$22:49:48.0  \\
FSR\,882&05:27:51.1&$+$16:53:49.2  \\
FSR\,884&($\dagger$)& ($\dagger$) \\
\hline
\multicolumn{10}{c}{Possible field fluctuations}\\
\hline
FSR\,744&($\dagger$)&($\dagger$)  \\
FSR\,776&06:07:24.0&$+$39:49:33.6  \\
FSR\,801&04:47:04.8&$+$24:54:00.0  \\
FSR\,841&05:06:13.4&$+$21:33:27.0  \\
FSR\,894&($\dagger$)&($\dagger$)  \\
FSR\,927&($\dagger$)&($\dagger$)  \\
FSR\,956&($\dagger$)&($\dagger$)  \\
\hline
\end{tabular}
\begin{list}{Table Notes.}
\item Cols.~2 and 3: Optimised central coordinates (Sect.~\ref{2mass}); $(\dagger)$ indicates
same central coordinates as in \citet{FSRcat}. Col.~4: Age, from 2MASS data. Col.~5:
$\rm\aV=3.1\,\ebv$. Col.~6: Distance from the Sun. Col.~7: \dgc\ calculated with $\Rgc=7.2$\,kpc
(\citealt{GCProp}) as the distance of the Sun to the Galactic centre. Cols.~8-10: Positional
components with respect to the Galactic plane.
\end{list}
\end{table*}

\section{Stellar radial density profiles}
\label{RDPs}

\begin{figure}
\resizebox{\hsize}{!}{\includegraphics{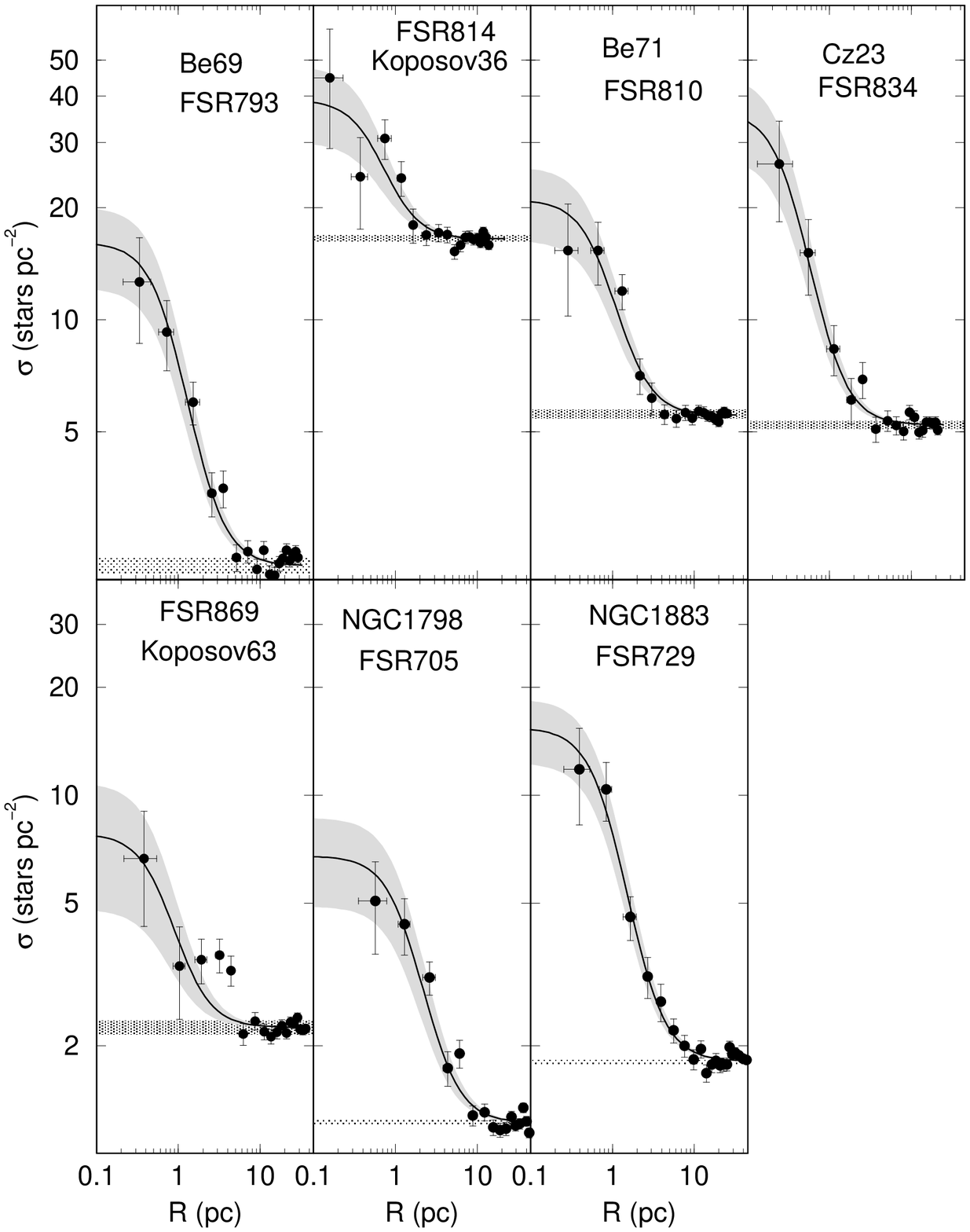}}
\caption{Stellar RDPs (filled circles) of confirmed open clusters built with photometry
that results from the use of colour-magnitude filters (Sect.~\ref{RDPs}). Solid line: best-fit
King profile. Horizontal shaded region: offset field stellar
background level. Gray regions: $1\sigma$ King fit uncertainty. Absolute scale is used.}
\label{fig7}
\end{figure}

\begin{figure}
\resizebox{\hsize}{!}{\includegraphics{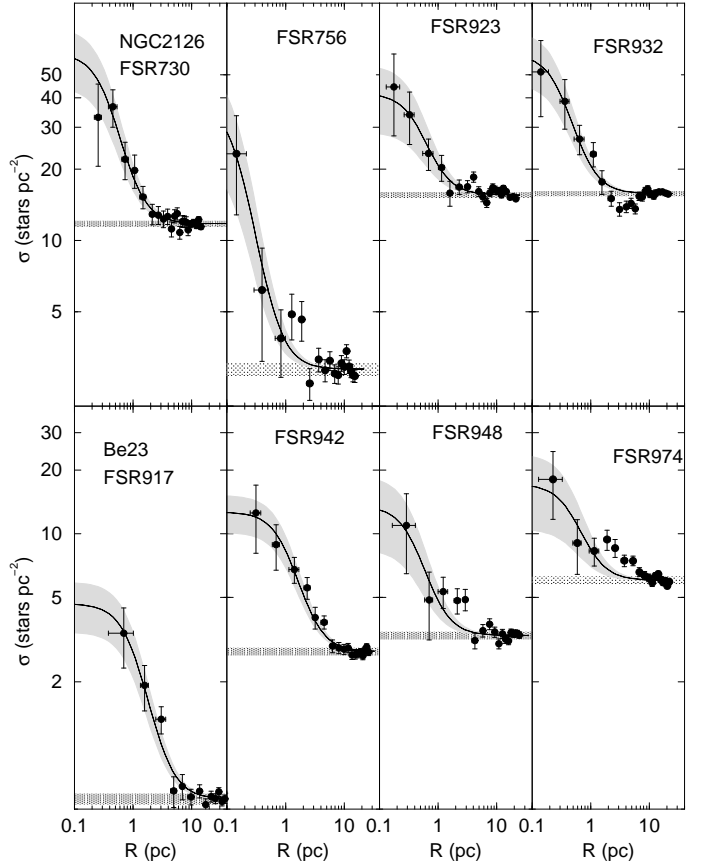}}
\caption{Same as Fig.~\ref{fig7} for the remaining OCs.}
\label{fig8}
\end{figure}

Before building the RDPs, we isolate the most probable cluster sequences with the colour-magnitude 
filters, which exclude the stars with colours different from those of the assumed cluster sequence.
Colour-magnitude filter widths are wide enough to include MS and evolved star distributions, and the
respective $1\sigma$ photometric uncertainties. They should also account for formation or dynamical
evolution-related effects, such as enhanced fractions of binaries (and other multiple systems) towards
the central parts of clusters, since such systems tend to widen the MS (e.g. \citealt{HT98};
\citealt{Kerber02}; \citealt{BB07}; \citealt{N188}). The filters for the present OCs are shown in the
bottom panels of Figs.~\ref{fig2} to \ref{fig5}. The contribution of residual field stars,
with similar colours to those of the cluster, to the RDPs is statistically quantified by means of  
comparison with the field. In practical terms, the use of colour-magnitude filters in cluster sequences 
enhances the contrast of the RDP with respect to the background (e.g. \citealt{BB07}). The corresponding
radial profiles of the 16 open clusters are given in Figs.~\ref{fig7} and \ref{fig8}.

\begin{figure}
\resizebox{\hsize}{!}{\includegraphics{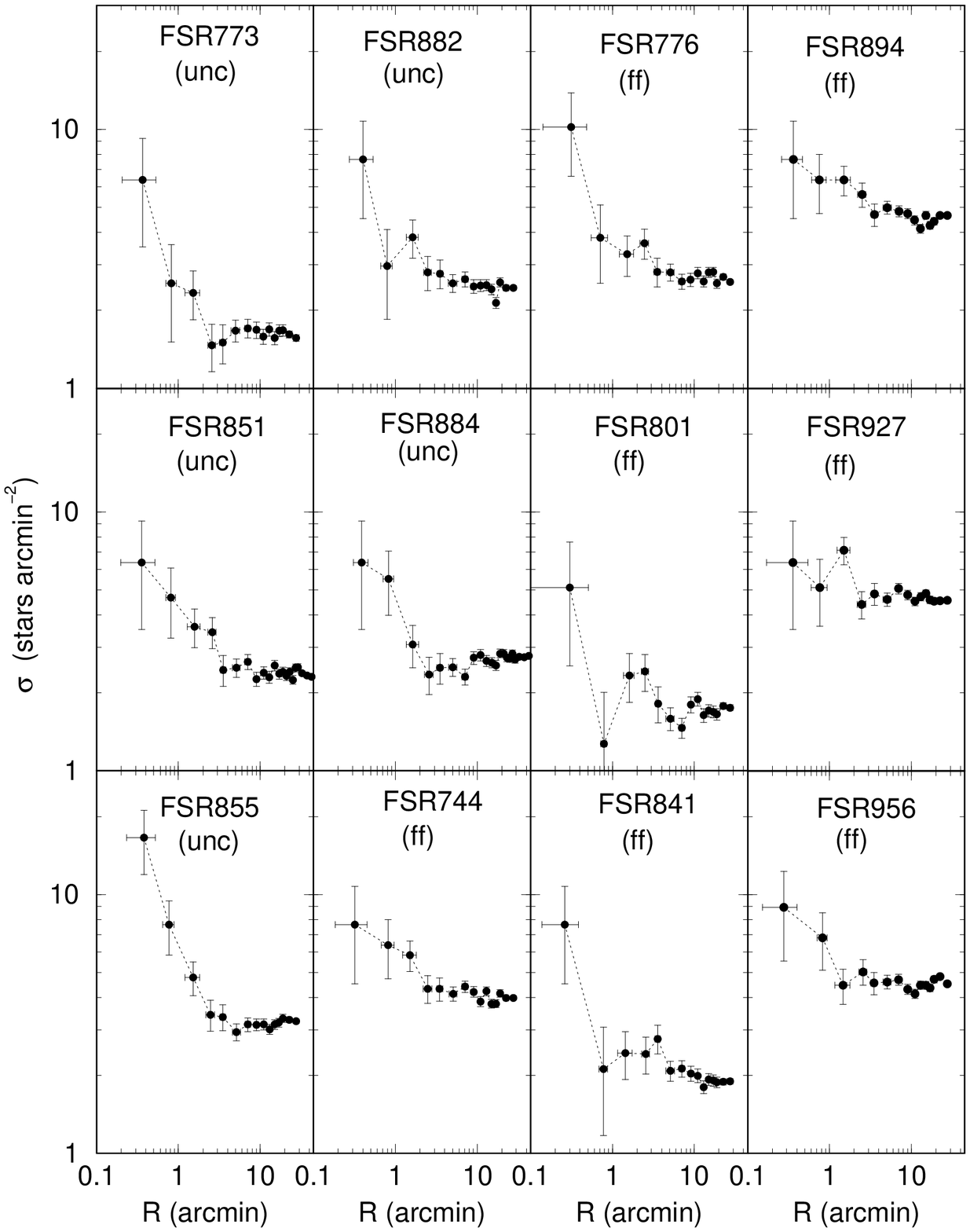}}
\caption{RDPs of the uncertain cases (unc) and the possible field fluctuations (ff), in angular 
units. In general, the RDPs of the possible field fluctuations are narrower and more irregular  
than those of the OCs (Figs.~\ref{fig7} and \ref{fig8}) and uncertain cases.}
\label{fig9}
\end{figure}

Star clusters, in general, have RDPs that can be described by a well-defined analytical profile,
characterised by parameters that are related to cluster structure. The most often used are the
single-mass, modified isothermal sphere of \citet{King66}, the modified isothermal sphere of
\citet{Wilson75}, and the power-law with a core of \citet{EFF87}. Because of the significant
error bars (Figs.~\ref{fig7} and \ref{fig8}), we use the analytical profile
$\sigma(R)=\sigma_{bg}+\sigma_0/(1+(R/R_C)^2)$, where $\sigma_{bg}$ is the residual background
density, $\sigma_0$ is the central density of stars, and \rc\ is the core radius. This function is 
similar to that by \cite{King1962} to describe the surface brightness profiles in the central parts
of globular clusters. $\sigma_0$ and the core radius (\rc) are derived from the RDP fit, while 
$\sigma_{bg}$ is measured in the respective comparison field. Because of the 2MASS photometric
depth, which in most cases corresponds to a cutoff for stars brighter than $\jj\approx16$, 
$\sigma_0$ should be taken as a lower limit to the actual central number-density.

Fit parameters are given in Table~\ref{tab6}, and the best-fit solutions are superimposed on 
the colour-magnitude filtered RDPs (Figs.~\ref{fig7} and \ref{fig8}). As expected, the adopted 
King-like function describes well the above RDPs over the full radial range, within uncertainties.
Table~\ref{tab6} also gives the density contrast parameter $\delta_c=1+\sigma_0/\sigma_{bg}$,
which is related to the difficulty of cluster detection against the background, and the RDP
radius \rl, which corresponds to the distance from the cluster centre where RDP and
background become statistically indistinguishable. Although most of the cluster stars are contained 
within $\rl$, it is smaller than the tidal radius. For instance, in populous and relatively high 
Galactic latitude OCs such as M\,67, NGC\,188, and NGC\,2477, the RDP radii are a factor
$\sim0.5 - 0.7$ of the respective tidal radii (\citealt{DetAnalOCs}). 

\begin{table*}
\caption[]{Structural parameters measured in the RDPs built with colour-magnitude filtered photometry}
\label{tab6}
\renewcommand{\tabcolsep}{2.95mm}
\renewcommand{\arraystretch}{1.4}
\begin{tabular}{lcccccccccc}
\hline\hline
&&&\multicolumn{7}{c}{RDP}\\
\cline{4-10}
Cluster&$1\arcmin$&&$\sigma_{bg}$&$\sigma_0$&$\delta_c$&\rc&\rl&\rc&\rl \\
       &(pc)&&$\rm(stars\,pc^{-2})$&$\rm(stars\,pc^{-2})$&&(pc)&(pc)&(\arcmin)&(\arcmin)\\
(1)&(2)&&(3)&(4)&(5)&(6)&(7)&(8)&(9)\\
\hline
Be\,23   &1.931&&$0.6\pm0.1$&$4.1\pm1.2$&$8.3\pm2.2$&$1.1\pm0.3$ &$10.0\pm2.0$ &$0.55\pm0.15$&$5.2\pm1.0$\\
Be\,69   &1.000&&$1.9\pm0.1$&$14.0\pm3.9$&$8.5\pm2.1$&$0.8\pm0.2$&$5.1\pm0.2$ &$0.80\pm0.16$&$5.1\pm0.2$\\
Be\,71   &0.862&&$5.5\pm0.1$&$15.5\pm4.7$&$3.8\pm0.8$&$0.8\pm0.2$&$4.9\pm0.2$ &$0.92\pm0.21$&$5.7\pm0.2$\\
Cz\,23   &0.728&&$5.2\pm0.1$&$31.1\pm9.0$&$7.0\pm1.7$&$0.4\pm0.1$&$3.6\pm0.1$ &$0.49\pm0.04$&$4.9\pm0.7$\\
NGC\,1798&1.731&&$1.2\pm0.1$&$5.6\pm1.9$&$5.5\pm1.5$&$1.4\pm0.4$ &$13.0\pm2.0$ &$0.82\pm0.21$&$7.5\pm1.1$\\
NGC\,1883&1.083&&$1.8\pm0.1$&$13.6\pm3.1$&$8.5\pm1.7$&$0.9\pm0.1$ &$10.0\pm1.0$ &$0.82\pm0.13$&$9.2\pm1.1$\\
NGC\,2126&0.298&&$11.8\pm0.2$&$49.6\pm17.0$&$5.2\pm1.4$&$0.4\pm0.1$ &$3.5\pm0.5$ &$1.34\pm0.33$&$11.7\pm1.7$\\
FSR\,756&0.521&&$2.9\pm0.1$&$36.0\pm17.0$&$13.5\pm5.9$&$0.2\pm0.1$ &$2.2\pm0.3$ &$0.31\pm0.19$&$4.2\pm0.6$\\
FSR\,814 &0.475&&$16.5\pm0.2$&$22.6\pm10.2$&$2.4\pm0.6$&$0.6\pm0.2$ &$2.5\pm0.2$ &$1.24\pm0.44$&$5.3\pm0.4$\\
FSR\,869 &1.232&&$2.3\pm0.1$&$5.6\pm2.9$&$3.5\pm1.3$&$0.7\pm0.3$ &$6.2\pm1.0$ &$0.54\pm0.27$&$5.0\pm0.2$\\
FSR\,923&0.447&&$15.6\pm0.2$&$26.1\pm13.0$&$2.7\pm0.8$&$0.5\pm0.2$ &$4.0\pm1.0$ &$1.12\pm0.42$&$8.9\pm0.5$\\
FSR\,932&0.444&&$15.8\pm0.2$&$45.1\pm15.2$&$3.9\pm1.0$&$0.4\pm0.1$ &$3.0\pm0.5$ &$0.79\pm0.20$&$6.7\pm1.1$\\
FSR\,942&0.900&&$2.8\pm0.1$&$9.9\pm2.5$&$4.6\pm0.9$&$1.1\pm0.2$ &$8.0\pm1.5$ &$1.25\pm0.24$&$8.9\pm1.7$ \\
FSR\,948&0.831&&$3.3\pm0.1$&$10.2\pm5.1$&$4.1\pm1.6$&$0.4\pm0.2$ &$4.2\pm0.5$ &$0.50\pm0.34$&$5.0\pm0.4$\\
FSR\,974&0.748&&$6.0\pm0.1$&$11.1\pm6.0$&$3.0\pm1.2$&$0.5\pm0.3$ &$9.0\pm1.0$ &$0.71\pm0.39$&$12.0\pm2.0$\\
\hline
\end{tabular}
\begin{list}{Table Notes.}
\item $(\dagger)$: FSR\,1644; $(\ddagger)$: FSR\,1723.
Col.~2: arcmin to parsec scale. To minimise degrees of freedom in RDP fits with the King-like
profile (see text), $\sigma_{bg}$ was kept fixed (measured in the respective comparison fields) while
$\sigma_0$ and \rc\ were allowed to vary. Col.~5: cluster/background density contrast
($\delta_c=1+\sigma_0/\sigma_{bg}$), measured in colour-magnitude filtered RDPs.
\end{list}
\end{table*}

The empirical determination of a cluster RDP radius depends on the relative levels of RDP and
background (and respective fluctuations). Thus, dynamical evolution may indirectly affect the
measurement of the RDP radius. This occurs because mass segregation drives preferentially low-mass
stars to the outer parts of clusters, which tends to lower the cluster/background contrast in these
regions as clusters age. As an observational consequence, lower values of the RDP radii are
expected to be measured, especially for clusters projected against dense fields. However, simulations
(\citealt{BB07}) of OCs with the structure described by a King-like profile, and projected against
different backgrounds, show that, provided not exceedingly high, background levels may produce RDP
radii underestimated by about 10--20\% with respect to the intrinsic values. The core radius, on the
other hand, is almost insensitive to background levels (\citealt{BB07}). This occurs because \rc\ is
derived from fitting the King-like profile to a distribution of RDP points, which minimises background
effects. 

The RDPs of the cases with uncertain CMD morphology and the possible field fluctuations are shown
in Fig.~\ref{fig9}. A narrow excess in the stellar RDPs near the centre is present in all cases, 
but they are quite different from a King-like profile, especially for the possible field fluctuations.

\section{Discussion}
\label{Disc}

Evidence gathered from the photometric (Sect.~\ref{CMDs}) and stellar radial distribution 
(Sect.~\ref{RDPs}) analyses indicate that, besides the 9 previously known OCs, 6 are newly 
identified ones, 6 are uncertain cases, while 7 are possible field fluctuations.

\subsection{Open clusters}
\label{ConfClus}

This group contains Be\,69, Be\,71, FSR\,814, Cz\,23, FSR\,869, NGC\,1798,
NGC\,1883, NGC\,2126, Be\,23, FSR\,756, FSR\,923, FSR\,932, FSR\,942, FSR\,948, and FSR\,974.
They have well-defined decontaminated CMD sequences (Figs.~\ref{fig2} - \ref{fig5}), relatively
high values of the parameter \ns\ (both in magnitude bins - Table~\ref{tab3}, and the integrated 
one - Table~\ref{tab4}), as well as King-like RDPs (Figs.~\ref{fig7} and \ref{fig8}). Astrophysical 
parameters (age, distance, reddening, core and RDP radii) could be measured for these clusters.
The average value of \ns\ is $\langle\ns\rangle=5.5\pm1.5$.

\subsection{Uncertain cases}
\label{DubCases}

In general, objects of the second group have less-defined decontaminated CMD sequences (Fig.~\ref{fig6}) 
than those in the first group, which is consistent with the lower-level of the integrated \ns, whose 
average is $\langle\ns\rangle=3.2\pm0.4$. The objects are FSR\,773, FSR\,851, FSR\,855, FSR\,882, FSR\,884, 
and FSR\,911. The 2MASS CMD sequences of this group cannot be used to unambiguously classify them as
star clusters. However, the relatively broad RDPs, with a high central surface density (Fig.~\ref{fig9}),
are rather similar to the RDPs of typical star clusters (e.g. Figs.~\ref{fig7} and \ref{fig8}).

Although the decontaminated CMDs and RDPs taken together might suggest star clusters of different 
ages, deeper observations are required to probe the existence of the TO and MS. Deeper photometry 
is essential in most cases, especially for old and/or distant OCs for which the TO is close to the 
2MASS limiting magnitude.

With respect to FSR\,911, in \citet{Boch1} we concluded that it is not the same objects as Bochum\,1, 
but rather an uncertain object, since there are other small-scale stellar concentrations in the area.

\subsection{Possible field fluctuations}
\label{FieldFluc}

Decontaminated CMDs of this group (Fig.~\ref{fig6}) have \ns-values significantly lower than those of
the star clusters (Sect.~\ref{ConfClus}) but of the same order as the uncertain cases (Sect.~\ref{DubCases}),
with $\langle\ns\rangle=3.0\pm0.7$. The fact that they have $\ns\sim3$ is consistent with the method
employed by \citet{FSRcat} to detect overdensities. However, in most cases the RDP excess is very
narrow and irregular, restricted to the first bins, quite different from a King-like profile (e.g.
Fig.~\ref{fig6}). In this group are FSR\,744, FSR\,776, FSR\,801, FSR\,841, FSR\,894, FSR\,927, and
FSR\,956.

Objects in this group have essentially featureless CMDs, and RDPs with important deviations from
cluster-like profiles. They appear to be $\sim3\sigma$ fluctuations of the stellar field over which
they are projected. Deep observations are also important to further probe the nature of these
overdensities.

\subsection{Comparison with previous studies}
\label{ComPrev}

\begin{figure}
\resizebox{\hsize}{!}{\includegraphics{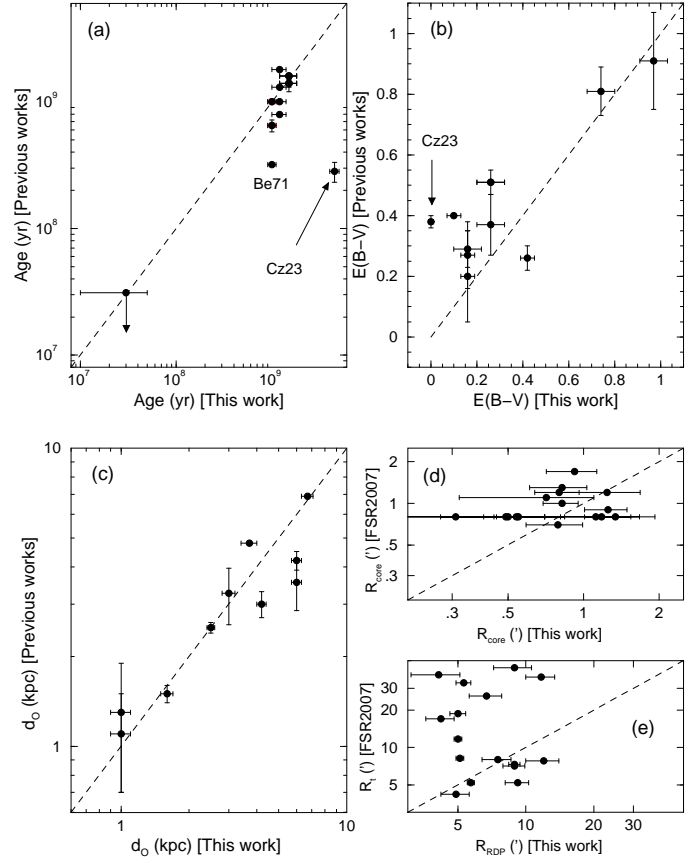}}
\caption{Presently derived parameters for the OCs are compared to those in previous studies. Panels (d)
and (e): the core and RDP radii derived here are compared to those in common with \citet{FSRcat};
we note that the latter authors use the tidal radii. Identity is indicated by the dashed line. }
\label{fig10}
\end{figure}

In Fig.~\ref{fig10} we compare parameters derived in this work for the confirmed OCs with those
found in studies of other authors (see the references given in Sect.~\ref{targets} for details on
literature parameters of each OC).  All the available data are used in the comparisons but,
since some parameters have similar values in different works, the near-coincidences cannot be 
individually distinguished in Fig.~\ref{fig10}. The upper limit on the age of FSR\,814 (\citealt{GKZ07})
is shown by an arrow. Parameters of Cz\,23 were derived by us in 
\citet{Cz23} with 2MASS photometry and the same methods as in the present paper, and these are 
taken for comparison with other works. Age (panel a), reddening (b), and distance from the Sun (c) present a good agreement, except for
the age and reddening of Cz\,23. We note that the differences in Cz\,23 parameters with respect to
\citet{GKZ07} arise from the fact that their photometry is not decontaminated, which is essential for
this object (as discussed in \citealt{Cz23}). 

Finally, in panels (d) and (e) we compare the present values of the core and RDP radii with those in
\citet{FSRcat}\footnote{Actually, \citet{FSRcat} present core and tidal radii which, in populous clusters
is a factor $\sim2$ larger than the RDP radii (\citealt{DetAnalOCs}).}. Most of the
lack of correlation can be explained by the very different surface-density distributions that follow
from the observed and decontaminated photometries (e.g. Fig.~\ref{fig1}). Decontaminated surface densities,
in general, enhance the central overdensity with respect to the surroundings, especially for OCs. Thus, 
King profile fits to decontaminated data should produce smaller radii, as compared to the observed photometry.
However, we note that part of the discrepancy can be accounted for by the different fit methods
employed to derive the structural parameters. While we work with the 2-parameter King profile
(Sect.~\ref{RDPs}), which essentially describes the central parts of clusters, \citet{FSRcat} use the
complete 3-parameter law (\citealt{King66}) to derive the core and tidal radii.

\subsection{Relations among astrophysical parameters}
\label{Relat}

\begin{figure}
\resizebox{\hsize}{!}{\includegraphics{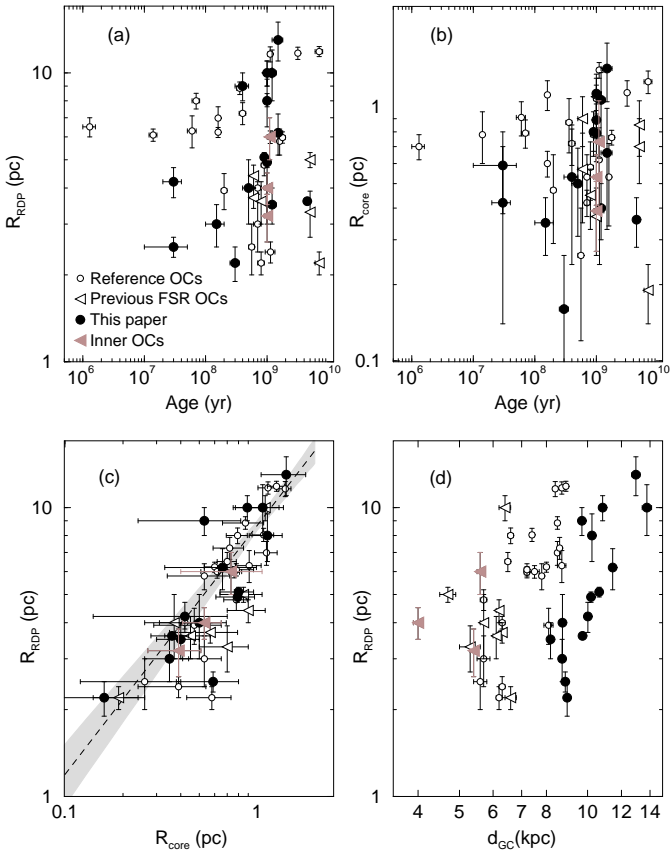}}
\caption{Relations involving astrophysical parameters of OCs. Empty circles: nearby OCs of the
reference sample. For comparison, similar FSR OCs from previous papers are shown as well
(triangles). The centrally-projected OCs FSR\,1744, FSR\,89 and FSR\,31 are shown as filled triangles.
Filled circles: the OCs dealt with in this work.}
\label{fig11}
\end{figure}

Diagrams with astrophysical parameters that can be used to investigate properties of OCs in different
environments have been presented in \citet{DetAnalOCs}. In this work we use them to investigate dependences
of core and RDP radii on cluster age and Galactocentric distance (Fig.~\ref{fig11}). As reference we
use {\em (i)}
a sample of bright nearby OCs (\citealt{DetAnalOCs}), including the two young ones NGC\,6611 (\citealt{N6611})
and NGC\,4755 (\citealt{N4755}), {\em (ii)} OCs projected against the central parts of the Galaxy
(\citealt{BB07}), and {\em (iii)} the recently identified OCs FSR\,1744, FSR\,89 and FSR\,31 (\citealt{OldOCs}),
which are also projected against the central parts of the Galaxy. OCs in sample {\em (i)} have ages in the
range $\rm70\,Myr\la age\la7\,Gyr$ and Galactocentric distances within $\rm5.8\la\dgc(kpc)\la8.1$. NGC\,6611
has $\rm age\approx1.3$\,Myr and $\dgc=5.5$\,kpc, and NGC\,4755 has $\rm age\approx14$\,Myr and $\dgc=6.4$\,kpc.
Sample {\em (ii)} OCs are characterised by $\rm600\,Myr\la age\la1.3\,Gyr$ and $\rm5.6\la\dgc(kpc)\la6.3$.
FSR\,1744, FSR\,89 and FSR\,31 are Gyr-class OCs at $\rm4.0\la\dgc(kpc)\la5.6$.

\begin{figure}
\resizebox{\hsize}{!}{\includegraphics{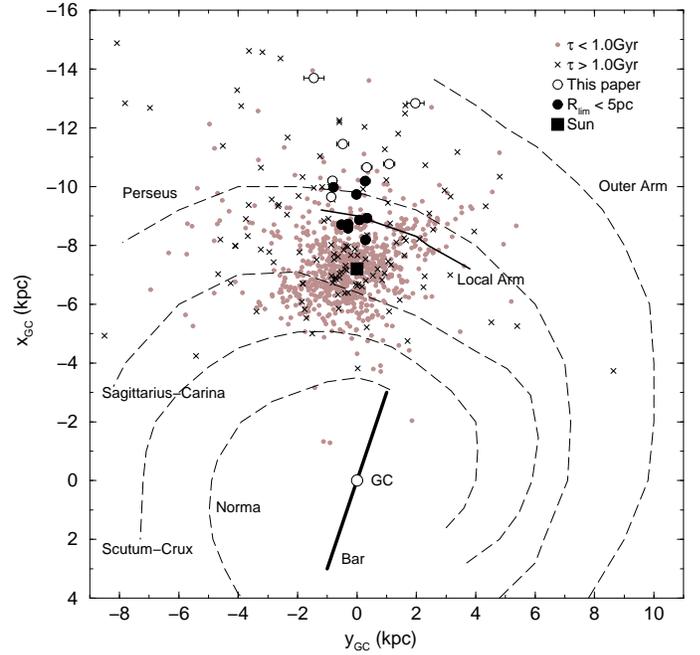}}
\caption{Spatial distribution of the present star clusters (circles) compared to the
WEBDA OCs with ages younger (gray circles) and older than 1.0\,Gyr (`x'). Small FSR OCs,
i.e. those with $\rl<5$\,pc, are shown as filled circles. Clusters are overplotted on a 
schematic projection of the Galaxy as seen from the North pole, with 7.2\,kpc as the Sun's 
distance to the Galactic centre. Main structures are identified.}
\label{fig12}
\end{figure}

The dependence of RDP and core radii on cluster age, which is intimately related to cluster
survival/dissociation rates, is examined in panels (a) and (b), respectively. Most of the present FSR OCs
appear to have both kinds of radii smaller than those of OCs of similar age. Besides, an accumulation of
small-radii (especially RDP radius) OCs appears to occur at $\sim0.5-1$\,Gyr, the typical time-scale of
OC disruption processes near the Solar circle (e.g. \citealt{Bergond2001}; \citealt{Lamers05}).

For most of the OCs in samples {\em (i)} and {\em (ii)}, the core and RDP radii follow the relation
$\rl=(8.9\pm0.3)\times R_{\rm core}^{(1.0\pm0.1)}$ (panel c), which implies a similar scaling in both
kinds of radii, in the sense that on average, larger clusters tend to have larger cores, at least for
$\rm 0.5\la\rc(pc)\la1.5$ and $\rm 5\la\rl(pc)\la15$. Similar relations between core and RDP radii were
also found by \citet{Nilakshi02}, \citet{Sharma06}, and \citet{MacNie07}.

A first-order dependence of OC size on Galactocentric distance is suggested by panel (d), as previously
discussed by \citet{Lynga82} and \citet{Tad2002}. The approximately linear relation between core and RDP
radii (panel c) implies a similar dependence of \rc\ with \dgc. Part of this relation may be primordial, in
the sense that the higher density of molecular gas in central Galactic regions may have produced clusters with
smaller core radii, as suggested by \citet{vdBMP91} to explain the increase of GC radii with Galactocentric
distance. In addition, there is the possibility that the core size may also be a function of the binary fraction
and its evolution with age, so that loss of stars may not be the only process to determine sizes. The present
FSR OCs also follow the trend of increasing RDP radii with Galactocentric distance, but in most cases the
values are lower than those of OCs at similar \dgc, especially for $\dgc\sim8-10$\,kpc.

Finally, in Fig.~\ref{fig12} we show the spatial distribution in the Galactic plane of the present FSR
OCs, compared to that of the OCs in the WEBDA database. For comparison purposes we consider two age
groups, clusters younger and older than 1\,Gyr. As expected, old OCs are found preferentially outside
the Solar circle, and the inner Galaxy contains few OCs so far detected.

\begin{figure}
\resizebox{\hsize}{!}{\includegraphics{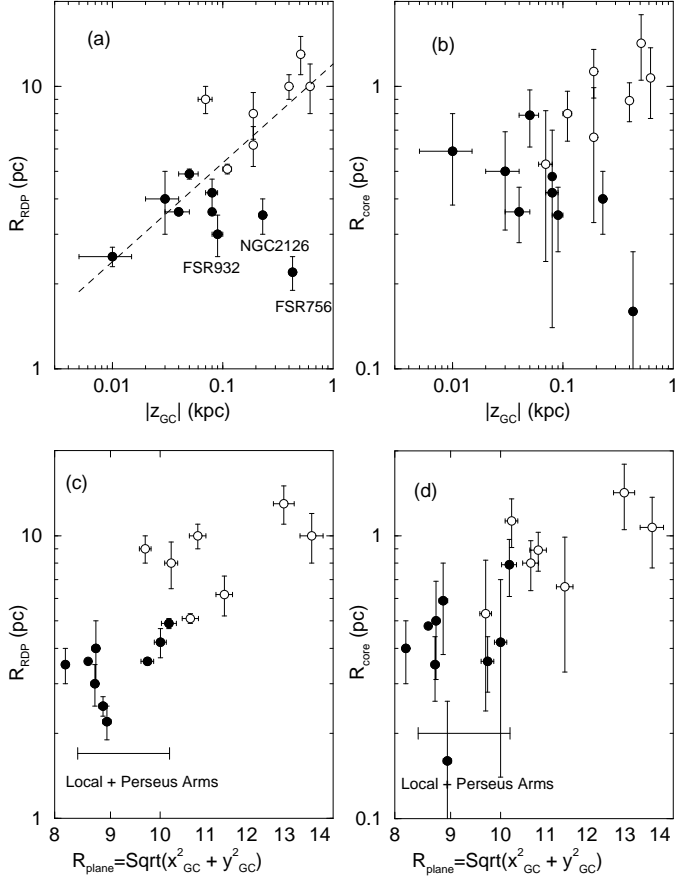}}
\caption{Cluster radii as a function of the vertical distance to the plane (top panels) and the on-plane
distance (bottom). Empty circles: OCs with RDP radius larger than 5\,pc. Filled circles: $\rl<5$\,pc.
The dashed line in panel (a) represents the fit $\rl=(12\pm2)\times|\zgc|^{0.35\pm0.06}$. FSR\,756, FSR\,932,
and NGC\,2126 are not included in the fit. The drop in RDP radii occurs in the region of the Local
$+$ Perseus Arms (panel c). }
\label{fig13}
\end{figure}

The spiral arm structure of the Milky Way (Fig.~\ref{fig12}) is based on \citet{Momany06} and
\citet{DrimSpe01}, as derived from H\,II regions, and molecular clouds (see e.g.
\citealt{Russeil03}). The Galactic bar is shown with an orientation of $14^\circ$ and
3\,kpc in length (\citealt{Freudenreich98}; \citealt{Vallee05}).

Despite the fact that the FSR OCs of the present sample are located outside the Solar circle,
some of them appear to present small core and RDP radii (Fig.~\ref{fig11}), as compared
to OCs at similar Galactocentric distances, which might suggest significant dynamical evolution.
This may be consistent with tidal stresses more important than previously assumed. Based on the
distribution of \rl\ with respect to \dgc\ (panel d of Fig.~\ref{fig11}), we take $\rl<5$\,pc 
as a representative value for the upper limit of the RDP radius of the smaller OCs outside the
Solar circle. We point out that the spatial distribution of the FSR OCs (Fig.~\ref{fig12}) suggests 
that such small OCs have interacted with the Perseus and Local arms, especially by means of encounters 
with giant molecular clouds. Molecular clouds more massive than $\sim10^6\,\ms$ are found in the
Solar neighbourhood (e.g. \citealt{Solo87}). Collision with such structures is another potential 
mechanism to decrease cluster mass, especially for low-mass clusters (e.g. \citealt{Wielen71}; 
\citealt{Wielen91}; \citealt{Gieles06}; \citealt{GAP07}). 

This point can be further investigated by examining the dependence of the core and RDP radii
with cluster position in the Galaxy (Fig.~\ref{fig13}). With respect to the vertical distance to
the Galactic plane $|\zgc|$, the present FSR OCs are distributed up to $|\zgc|\approx0.6$\,kpc
(Table~\ref{tab5}). When plotted as a function of distance to the Galactic plane, the RDP
radii present a general trend of increase with $|\zgc|$ (panel a). The exceptions are FSR\,756,
FSR\,932, and NGC\,2126. These OCs excluded, the distribution of the remaining objects follows
relatively tightly the power-law $\rl=(12\pm2)\times|\zgc|^{0.35\pm0.06}$. Such a relation is
consistent with a lower-frequency of encounters with giant molecular clouds, and with the disk,
for OCs at high $|\zgc|$ with respect to those orbiting closer to the plane. Although with more 
scatter, the core radii distribution (panel b) is similar to that of the RDP radii, which
again suggests dynamical effects. A similar dependence of OC radius with $|\zgc|$ was 
observed by \citet{Tad2002}, who also noted that, in the Solar neighbourhood, OCs with radius 
smaller than 5\,pc are found preferentially close to the plane. We remark that part of this effect
can be accounted for by the dependence of completeness on \zgc\ (\citealt{DiskProp}). Since the 
average background$+$foreground contamination decreases for high \zgc, the external parts of an
OC (i.e. with intrinsically low surface brightness) can be detected for a larger distance than 
for a low-\zgc\ objects.

As shown in panels (c) and (d), when the on-plane distance $\left(R_{plane}=\sqrt{\xgc^2 + \ygc^2}\right)$
is considered, a trend of increasing radii with $R_{plane}$ occurs, similarly to the relation of \rl\
with \dgc\ (Fig.~\ref{fig11}). However, there appears to occur a gap around the region corresponding
to the Local and Perseus arms. Besides, the position of the FSR OCs with $\rl<5$\,pc roughly coincides
with the inter-arm region (Figs.~\ref{fig12} and \ref{fig13}).

As a caveat we note that the above arguments are based on a relatively small number of OCs.

\section{Concluding remarks}
\label{Conclu}

The identification and derivation of astrophysical parameters of new star clusters in the Galaxy
provide valuable inputs and constraints to studies of star formation and evolutionary processes, 
dynamics of N-body systems, cluster disruption time scales, the geometry of the Galaxy, among 
others.

For this work we selected all star cluster candidates projected nearly towards the anti-centre
($160^\circ\le\ell\le200^\circ$ and $-20^\circ\le b\le20^\circ$), from the catalogue of \citet{FSRcat}.
Identified as stellar overdensities by \citet{FSRcat}, the candidates were classified by them as probable and
possible star clusters, with quality flag '0' or '1'. The 28 such candidates were analysed by means of
2MASS field-star decontaminated colour-magnitude diagrams, colour-magnitude filters, and stellar radial
density profiles.

Of the 28 candidates, 15 have cluster-like CMD morphologies and King-like RDPs. Among these are the 
previously catalogued open clusters NGC\,1798, NGC\,1883, NGC\,2126, Be\,23, Be\,69, Be\,71, and Cz\,23, 
and 2 that have been recently identified as OCs by \citet{GKZ07} in a similar search for stellar 
overdensities (FSR\,814 and FSR\,869). In this paper, the remaining 6 candidates are shown to be OCs, 
and their properties are investigated. These star clusters have ages in the range $\sim4$\,Myr
to $\sim4.5$\,Gyr, distances from the Sun within $\rm1.0\la\ds(kpc)\la6.7$, and Galactocentric distances
within $\rm8.2\la\dgc(kpc)\la13.8$. Six other candidates have CMDs and/or RDPs that suggest star clusters
of different ages, but deeper photometry is required to establish their nature. The remaining 7 overdensities
are probably fluctuations of the stellar field.

From the above numbers we estimate that the fraction of FSR anti-centre overdensities that turn out 
to be star clusters ($f_{SC}$) can be put in the range $54\%\la f_{SC}\la75\%$. The lower limit agrees 
with the $f_{SC}\approx50\%$ estimated by \citet{FSRcat} for all directions. Considering the 
anti-centre solid angle sampled in this paper, and the newly identified OCs (Sect.~\ref{ConfClus}), 
the density of these OCs is $\rm\eta_{AC}\sim74\,sterad^{-1}$. Similar arguments applied
to a central direction (e.g. \citealt{ProbFSR}) yields a density $\rm\eta_{C}\sim14\,sterad^{-1}$.
Thus, the density of new OCs in the anti-centre is a factor of $\sim5.3$ higher than towards
the centre. This is probably associated to the high level of the Galactic stellar field
in central directions. Significant statistical fluctuations are expected to occur in such  
fields, in scales similar to those produced by star clusters (e.g. \citealt{ProbFSR}).

WEBDA contains 75 OCs with known age and distance from the sun located in the same sector as
that probed in this work. Thus, the accurate parameters derived here represent an increase of 
$\approx11\%$ to the database.

A considerable fraction of the FSR OCs studied here are located close to the Local and Perseus Arms. 
These OCs appear to be abnormally small, which probably can be accounted for by mass-loss due to 
dynamical interaction with giant molecular clouds and the spiral arms.

Catalogues of star cluster candidates, coupled with field-star decontamination and stellar radial 
profiles, have become a powerful tool to detect and derive astrophysical parameters of new star 
clusters in the Galaxy. A consequence of this kind of study will be a better definition of the OC
parameter space, with reflexes on the understanding of the star-formation rate, star-cluster dynamical 
evolution, and Galaxy evolution and structure.

\section*{Acknowledgements}
We thank the anonymous referee  for suggestions.
This publication makes use of data products from the Two Micron All Sky Survey, which is a 
joint project of the University of Massachusetts and the Infrared Processing and Analysis
Center/California Institute of Technology, funded by the National Aeronautics and Space 
Administration and the National Science Foundation. This research has made use of the WEBDA 
database, operated at the Institute for Astronomy of the University of Vienna. We acknowledge 
partial support from CNPq (Brazil). 


\end{document}